\documentclass[12pt]{article}
\usepackage{graphicx}

\newcommand\pubnumber{SNSN-323-63}
\newcommand\pubdate{\today}

\def\hawaii{Department of Physics\\University of Hawaii at Manoa, Honolulu - HI (US)}

\def\Title#1{\begin{center} {\Large #1 } \end{center}}
\def\Author#1{\begin{center}{ \sc #1} \end{center}}
\def\Address#1{\begin{center}{ \it #1} \end{center}}

\newcommand\pubblock{\rightline{\begin{tabular}{l} \pubnumber\\
         \pubdate  \end{tabular}}}
\newenvironment{Abstract}{\begin{quotation}  }{\end{quotation}}
\newenvironment{Presented}{\begin{quotation} \begin{center}
             PRESENTED AT\end{center}\bigskip
      \begin{center}\begin{large}}{\end{large}\end{center} \end{quotation}}

\def\beq{\begin{equation}}
\def\eeq#1{\label{#1}\end{equation}}
\def\eeqn{\end{equation}}

\def\beqa{\begin{eqnarray}}
\def\eeqa#1{\label{#1}\end{eqnarray}}
\def\eeqan{\end{eqnarray}}

\let\bar=\overbar

\def\Dslash{\not{\hbox{\kern-4pt $D$}}}
\def\dslash{\not{\hbox{\kern-2pt $\del$}}}

\def\msb{{\bar{\ssstyle M \kern -1pt S}}}

\begin{document}
\begin{titlepage}
\pubblock

\vfill
\Title{Primary Cosmic Ray Proton Flux Measured by AMS-02}
\vfill
\Author{ C. Consolandi, on Behalf of the AMS-02 Collaboration}
\Address{\hawaii}
\vfill
\begin{Abstract}
The Alpha Magnetic Spectrometer (AMS-02) is a high energy particle detector designed to study origin and nature of cosmic rays up to a few TV from space. It was installed on the International Space Station (ISS) on May 19$^{th}$, 2011.
During the first two years of operation AMS-02
performed precise measurements of the proton flux. In the low rigidity range, from 1 GV to 20 GV, the proton flux was daily measured with a statistical error less than 1\%. In the same rigidity range a gradual decrease due to Solar modulation effect and transit variations due to Solar Flares and Coronal Mass Ejection were also observed.
In the rigidity range from 20 GV up to 100 GV instead, AMS-02 data show no drastic variation and the results are consistent with other experiments. Above 100 GV, AMS-02 proton flux exhibits a single power low behavior with no fine structures nor brakes.
\end{Abstract}

\vfill
\begin{Presented}
The 10th International Symposium on Cosmology and Particle Astrophysics (CosPA2013)\\
Honolulu, Hawai'i,  November 12$^{th}$--15$^{th}$, 2013
\end{Presented}
\vfill
\end{titlepage}
\def\thefootnote{\fnsymbol{footnote}}
\setcounter{footnote}{0}

\section{Proton Flux Introduction}

Protons constitute the most abundant part of the galactic cosmic rays. Knowing their absolute flux and their spectral shape is of fundamental importance for understanding the origin and propagation of galactic cosmic rays in our Galaxy. In addition, primary protons coming from the Sun are an important probe to study solar phenomena.
Proton spectral shape above 100 GeV has recently been under strong interest for the space particle community.
In this paper the accurate determination of the proton flux measured by AMS-02 is presented~\cite{bib:sada}.

\section{AMS-02 Detector}

AMS-02 was installed on the International Space Station (ISS) on May 19$^{th}$, 2011 to perform a unique and long duration mission ($\sim$20 years) in particle space physic research.
AMS-02~\cite{bib:amspf} is a magnetic spectrometer consisting
of several detectors~\cite{bib:amspf}. To perform proton flux analysis the following sub-detectors were involved:
the permanent magnet of 1.4~kG~\cite{bib:magnim}, the nine layers of the silicon Tracker with a maximum track path length of 3~m~\cite{bib:trnim}, the 4 layers of the time of flight system (TOF) which determines particles direction, velocity and constitutes the AMS-02 trigger~\cite{bib:tofnim}.
The particle rigidity, which is defined as the momentum over the charge, is obtained by means of a fitting procedure witch uses information of the three dimensional trajectory reconstructed by the tracker and bent by the magnetic field. The AMS-02 maximum detectable rigidity (MDR) for protons~\cite{bib:trrec,bib:tralg} is about 2~TV.

\section{Data Sample and Exposure Time}

Two years of AMS-02 data collected from May 19$^{th}$, 2011 to May 19$^{th}$, 2013 are analyzed in this work.
During each of the 6.3$\times10^7$ seconds of data taking the  AMS-02 live time was precisely measured.
The exposure time period is selected with second-by-second basis as follows:
\begin{itemize}
\item AMS-02 is required to be in the nominal data taking status,
\item the AMS-02 vertical axis must lie within 25$^\circ$ of the Earth zenith axis,
\item the measured rigidity is required to exceed by a factor 1.2 of the
maximal Stoermer cutoff\cite{bib:stoermer}.
\end{itemize}
The total exposure time varies according to the measured rigidity: at 1 GV it is equal to $1.52\times10^6$ seconds and rapidly increases with rigidity. For rigidities above 25 GV
it reaches the constant value of $5.12\times10^7$ seconds, which corresponds to an overall average live time fraction of 81.6\% for two years.

\section{Event Selection}
To reach the highest possible rigidity resolution,
events with at least one full span track in the tracker where selected. The full span track is defined to have hits in both outer most planes (layer 1 and 9 respectively).

\subsection*{Preselection}
The preselected sample is composed of events that are requested to have:
\begin{itemize}
\item velocity measured by at least three TOF layers consistent with a down going particle, and
\item linearly extrapolated trajectory of the TOF hit positions passing both tracker layer 1 and 9.
\end{itemize}

\subsection*{Proton Track Selection}
\begin{figure}[t]
 \centering
 \includegraphics[width=0.6\textwidth]{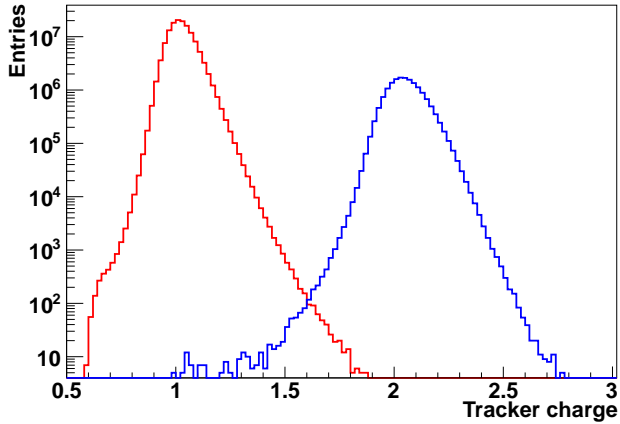} 
 \caption{Tracker charge distributions of proton -red curve- and Helium -blue curve- pure sample selected by an independent charge measurement by TOF.}
 \label{trchg}
\end{figure}

Proton candidates are selected among the preselected sample to have:
\begin{itemize}
\item at least one track reconstructed in the tracker with four planes inside the magnet bore, and
\item charge measured by the tracker consistent with $Z=1$ particle.
\end{itemize}
The tracker charge is determined by multiple measurements of energy loss in all tracker layers of the double sided silicon detector~\cite{bib:trchg}.
To estimate the $Z=1$ charge selection efficiency a pure proton sample, selected by an independent charge measurement performed with TOF~\cite{bib:tfchg}, was used. As shown in Fig.~\ref{trchg}, the $Z=1$ charge selection efficiency
is more than 99.9~\% over the whole rigidity range.


\subsection*{Final Sample Selection}
The final proton events, used for flux analysis, are required to have:
\begin{itemize}
\item at least one track with measured hit positions in two
coordinates both in layer 1 and 9, and
\item normalized $\chi^2$ of the fitting track, in the bending plane, less than 10.
\end{itemize}
After all these cuts $3.03\times10^8$ events were selected.

\subsection{Backgrounds}
Protons are the most abundant particles in primary cosmic rays.
However the possibility of backgrounds were studied and quantified.

As shown in Fig.~\ref{trchg} the probability that helium is mis-identified as proton is estimated to be less than 0.1~\% over the whole rigidity range. This estimation was performed with a pure helium sample selected with an independent TOF~\cite{bib:tfchg} charge measurement.

Pions that are produced in the atmosphere and in the material around AMS-02 can be a source of contamination for the proton sample at low energy. Nevertheless after the request of a measured rigidity to exceed the maximal Stoermer cutoff by a factor 1.2, the pion contamination was estimated to be less than 1~\% in the 1--2~GV region.

Electrons are mostly rejected by requesting a positive measured rigidity. However the possible misidentification of the charge sign, that may affect the high energy range of the spectra, is considered to be negligible because the spectral index of electron is steeper than the one of  protons~\cite{bib:amse}.

In this analysis, positrons were not separated from protons
but their total contribution is less than 1~\%.

Deuterons were not separated from protons as well,
but according to previous measurements~\cite{bib:bessdp}, the cosmic ray deuteron to proton ratio is 2$\sim$3~\% at 1~GV and it decreases with increasing rigidity.

\section{Flux Normalization}

Assuming an isotropic flux for rigidities greater than
the geomagnetic cutoff, the differential proton flux $J$
as a function of rigidity $R$ may be written as follows:
\begin{equation}
J(R)=\frac{N_{{\rm obs}}}{T_{{\rm exp.}}A_{{\rm eff.}}
  \varepsilon_{{\rm trg.}}\varepsilon_{{\rm trk.}}
	     {\rm d} R} \label{eqflx}
\end{equation}
where: $N_{\rm obs}$ is the number of selected events; $T_{\rm exp.}$ is the exposure time; $A_{\rm eff.}$ is the effective acceptance which includes both geometrical factor and efficiencies, without large migration of energy, due to hadronic interactions; $\varepsilon_{\rm trg.}$ is the trigger efficiency; $\varepsilon_{\rm trk.}$ is the selection efficiency of proton tracks, and ${\rm d}R$ is the rigidity bin width.

\subsection{Acceptance}
The effective acceptance is estimated by means of a
simulation technique~\cite{bib:sulvin}.
A dedicated program based on the GEANT-4.9.4 package~\cite{bib:geant4} was developed to simulate Monte Carlo events. In this program electromagnetic and hadronic interactions of particles in materials are simulated and AMS-02 detector response is generated as well. After the simulation process, the digitized signal undergoes to the same reconstruction as the one applied on data.
The acceptance, $A_{{\rm eff.}}$ is obtained as :
\begin{equation}
A_{{\rm eff.}}=A_{{\rm gen.}}\times\frac{N_{{\rm acc.}}}{N_{{\rm gen.}}}
\end{equation}
where: $A_{{\rm gen.}}$ is a geometrical factor of the Monte Carlo generation plane; $N_{{\rm gen.}}$ is the number of generated events, and $N_{{\rm acc.}}$ is the number of events which passed the preselection.

The generation plane is a 3.9$\times$3.9 m$^2$ square surface
on top of AMS-02 ($z=195$~cm), which corresponds to $A_{{\rm gen.}}=$ 47.8~m$^2$sr.
The obtained acceptance depends slightly (less than 5~\%)
on rigidity below 10~GV and is constant above 10~GV.
A systematic error of 2.8~\% due to the uncertainty
of energy dependence of the hadronic interaction probability is taken into account.

\subsection{Trigger efficiency}
\begin{figure}[t]
 \centering
 \includegraphics[width=0.6\textwidth]{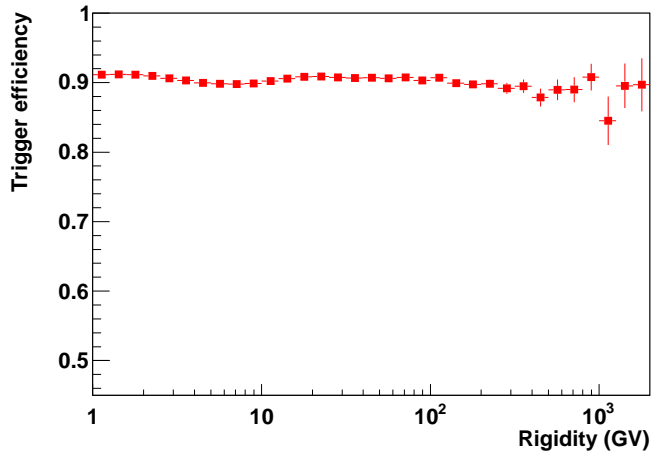} 
 \caption{Trigger efficiency as a function of measured rigidity.}
 \label{trgef}
\end{figure}

In the AMS-02 trigger logic different physics trigger conditions are implemented. The trigger logic was designed to maximize the efficiency for each particle species and to keep a sustainable rate of the recorded events.
In order to measure the trigger efficiency directly from data, 1/100 of the events with a coincidence of signals from at least 3 TOF planes are tagged as unbiased.
The trigger efficiency, $\varepsilon_{\rm trg.}$ is obtained as:
\begin{equation}
\varepsilon_{{\rm trg.}}=\frac{N_{{\rm phys.}}}
	   {N_{{\rm phys.}}+100\times N_{{\rm unb.}}}
\end{equation}
where: $N_{{\rm phys.}}$ is the number of events that passed the proton selection and triggered with any of the physics trigger conditions, and $N_{{\rm unb.}}$ is the number of events which passed the proton selection and are triggered as unbiased sample.
Fig.~\ref{trgef} shows the trigger efficiency as a function of measured rigidity. It is constant above 20~GV within 1~\%. The systematic error of 1~\% is due to the limited statistics of the unbiased sample.

\subsection{Track Reconstruction Efficiency}

\begin{figure}[t]
 \centering
 \includegraphics[width=0.6\textwidth]{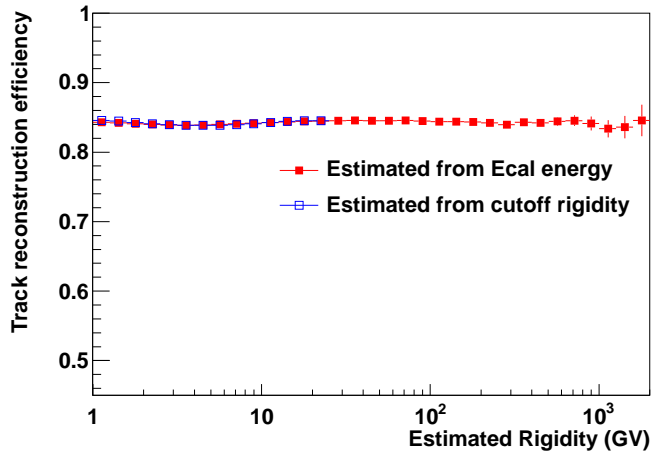} 
 \caption{Track reconstruction efficiency estimated from the
   preselection sample and TOF charge selection, which includes the events passing out of the tracker sensitive area (about 91~\%).
   Red squares: efficiency as a function of
   rigidity estimated by ECAL energy deposition. Blue open squares: efficiency as a function of cutoff rigidity.}
 \label{treff}
\end{figure}

The track reconstruction efficiency is estimated as the ratio of the number of events after the proton track selection over the number of events that passed the preselection and the independent TOF charge selection.
This last sample includes events passing out of the tracker sensitive area which is about 91~\%.
Fig.~\ref{treff} shows the track reconstruction efficiency as
a function of rigidity that was estimated by the energy deposition in the Electro-magnetic calorimeter (ECAL). This measure is consistent with the rigidity
estimated from the geomagnetic cutoff. The efficiency is constant within 1~\% over the whole energy range. The systematic error of 1~\% is due to the uncertainty of energy dependence.

\subsection{Efficiency Stability}
\begin{figure}
 \centering
 \includegraphics[width=0.6\textwidth]{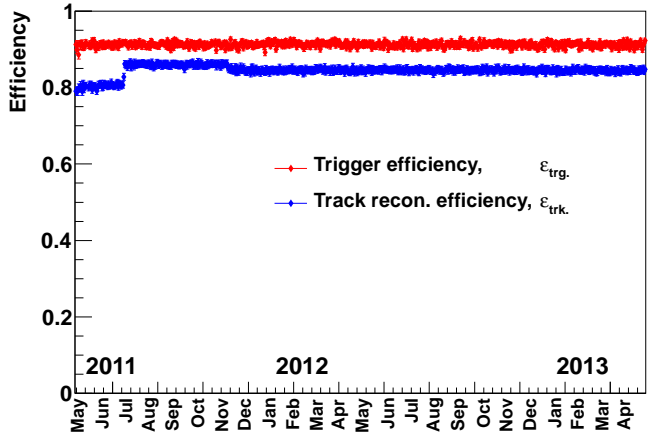} 
 \caption{Daily variation of
   the trigger efficiency ($\varepsilon_{{\rm trg.}}$), and the
   proton track efficiency ($\varepsilon_{{\rm trk.}}$)
   for rigidities above 20 GV.
 }
 \label{eftim}
\end{figure}

In Fig.~\ref{eftim}, the daily variation of efficiencies
$\varepsilon_{{\rm trg.}}$, and $\varepsilon_{{\rm trk.}}$
at rigidities greater than 20 GV are showed for the two years of data:
$\varepsilon_{{\rm trg.}}$ is constant within the statistical error of 0.7~\%.
For $\varepsilon_{{\rm trk.}}$
the small increase on July 24$^{th}$, 2011 is due to the improvement of the tracker calibration and the small drop on December 1$^{st}$, 2011 is due to a loss of 3~\% of tracker readout channels~\cite{bib:trop}.
All lost channels are in non-bending coordinate so the impact
on the rigidity measurement is negligible.

\section{Binning and Unfolding}
\label{ufold}
AMS-02 rigidity binning is chosen according to the resolution estimated with Monte Carlo simulation~\cite{bib:trrec}.
Due to the finite spectrometer resolution, the normalized rigidity distribution of selected protons was corrected for bin-to-bin migration effects.
The migration matrix was obtained with Monte Carlo simulation and parametrized with two Gaussians. The incident differential proton flux was obtained by means of an unfolding procedure based on the Bayes' theorem~\cite{bib:unfold} applied to the measured proton flux.
The unfolding errors were estimated by changing of about 10~\% the sigma of the resolution matrix. This 10~\% corresponds to our test beam data extrapolation error to that energy. In addition it was allowed a shift up to 1/20~TV$^{-1}$ of the average inverse rigidity measurement, which corresponds to the current tracker alignment knowledge that was obtained by
using electron and positron samples~\cite{bib:trrec,bib:tralg}.

\section{Error Estimation}
AMS-02 statistical errors are always less than 1~\% in the whole energy range. The Systematic error on the flux normalization, $\sigma_{{\rm norm.}}$, instead, is estimated as :
\begin{equation}
\sigma_{{\rm norm.}}=\sqrt{\sigma_{{\rm acc.}}^2+\sigma_{{\rm trg.}}^2+
                           \sigma_{{\rm trk.}}^2}=3.1\textrm{ \%}
\end{equation}
where $\sigma_{{\rm acc.}}= 2.8$~\% is the error on the acceptance estimation $\sigma_{{\rm trg.}}= 1.0$~\% is the error on the trigger efficiency and $\sigma_{{\rm trk.}}= 1.0$~\% is the error on the proton track efficiency.
As discussed in Section~\ref{ufold}, the systematic error due to the unfolding procedure, $\sigma_{{\rm unfold.}}$ is estimated
by changing the parametrization of the migration matrix and
it is less than 1~\% below 100 GV and 5.4~\% at 1 TV.

The total systematic error is obtained as the quadratic sum of
$\sigma_{{\rm norm.}}$ and $\sigma_{{\rm unfold.}}$ and it is 3.2~\% below 100~GV and 6.3~\% at 1~TV.

\section{Daily Flux Variation}

\begin{figure*}[t]
 \centering
 \includegraphics[width=0.9\textwidth]{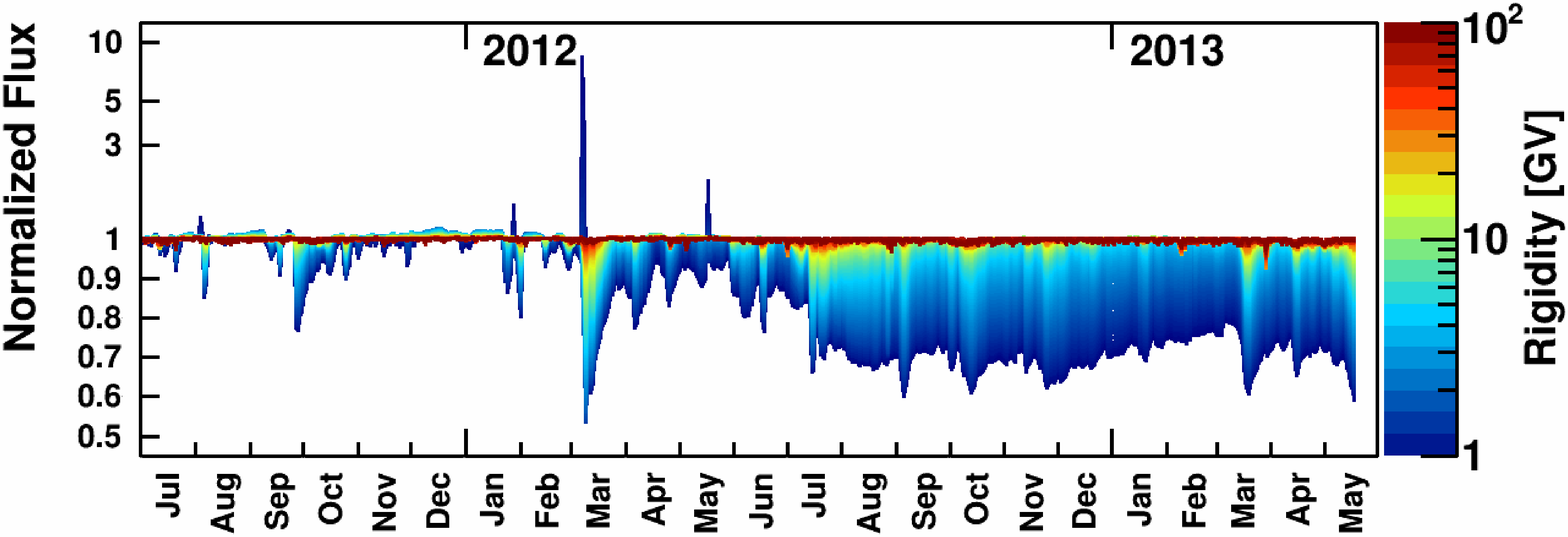} 
 \caption{Daily variation of the  normalized flux. The gradual decrease of the flux in the low rigidity region ($R<\sim$10 GV) was observed as well as some spikes in $\sim$1~GV which correspond to solar events on August 9$^{th}$, 2011 (X6.9), January 27$^{th}$, 2012 (X1.7), March 7$^{th}$, 2012 (X5.4 and X1.3), and May 17$^{th}$, 2012 (M5.1).}
 \label{fltim}
\end{figure*}

The AMS-02 proton flux is daily measured with a statistical error of $\sim$1~\% up to $\sim$20~GV.
Below 30~GV  it is affected by the solar activity as can be seen in Fig.~\ref{fltim} which shows the two years time variation of the normalized flux with increasing rigidity bins between 1 and 100 GV, from blue to red. Fluxes at the beginning of the mission are normalized to 1. As time passes by, it is possible to appreciate the gradual decrease of the flux in the low rigidity region ($R<\sim$10 GV) due to the solar modulation effect.
The large spike observed on March 7$^{th}$, 2012 corresponds to the \textit{March Solar Event} (X5.4 and X1.3 class solar flares and two fast Coronal Mass Ejections) which was the strongest of Solar Cycle 24$^{th}$. After the \textit{March Solar Event} AMS-02 has also observed the large Forbush decrease up to 30 GV which lasted for about three weeks.
Another spike on May 17$^{th}$, 2012 corresponds to the M5.1 solar flare which produced the first Ground Level Enhancement (GLE) of Solar Cycle 24$^{th}$.
Other small spikes in the plot correspond to further solar events on August 9$^{th}$, 2011 (X6.9) and January 27$^{th}$, 2012 (X1.7). Several other Forbush decreases, including the large one of September 27$^{th}$, 2011, are also visible.

\begin{figure}[!h]
 \centering
 \includegraphics[width=0.8\textwidth]{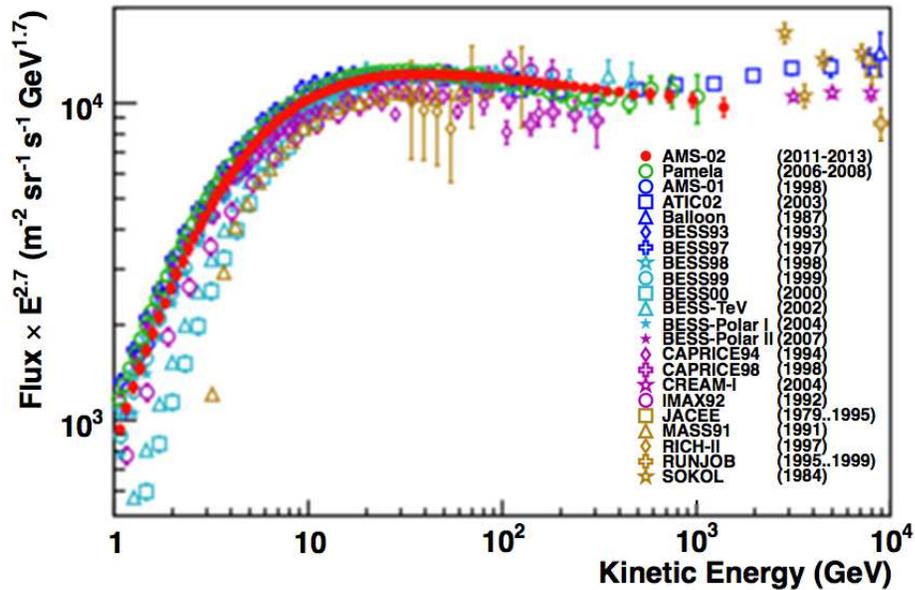} 
 \caption{The average proton flux over the two years of AMS-02 observation
   as a function of kinetic energy ($E$)
   multiplied by $E^{2.7}$ together with the previous experimental
   data~\cite{bib:pamela}--\cite{bib:crdb}.}
 \label{pfall}
\end{figure}

\section{Result and Conclusion}
Fig.~\ref{pfall} shows the average proton flux over the two years
of AMS-02 observation as a function of kinetic energy
multiplied by the corresponding bin central value~\cite{bib:bincen}
taken to the 2.7 power and compared with previous experimental
data~\cite{bib:pamela}--\cite{bib:crdb}.
In the high energy region above 100 GeV the spectrum
is consistent with a single power law spectra and
shows no fine structure nor break.


\begin{thebibliography}{99}

\bibitem{bib:sada}
S. Haino et al., ICRC (2013).

\bibitem{bib:amspf}
M. Aguilar et al., Phys. Rev. Lett 110 (2013) 141102.

\bibitem{bib:magnim}
K. Luebelsmeyer et al., Nucl. Instr. Meth. A 654 (2011) 639.

\bibitem{bib:trnim}
B. Alpat et al., Nucl. Instr. Meth. A 613 (2010) 207;

\bibitem{bib:tofnim}
A. Basili et al., Nucl. Instr. Meth. A 707 (2013) 99;
V. Bindi et al., Nucl. Instr. Meth. A 623 (2010) 968.

\bibitem{bib:trrec}
P. Zuccon et al., ICRC (2013) 1064.

\bibitem{bib:tralg}
C. Delgado et al., ICRC (2013) 1260.

\bibitem{bib:stoermer}
C. Stoermer, The Polar Aurora (Oxford University, London, 1950).

\bibitem{bib:trchg}
P. Saouter et al., ICRC (2013) 789.

\bibitem{bib:tfchg}
Q. Yan et al., ICRC (2013) 1097.

\bibitem{bib:amse}
S. Schael et al., ICRC (2013) 1257;
B. Bertucci et al., ICRC (2013) 1267.

\bibitem{bib:bessdp}
J.Z. Wang et al., Astrophs. J. 564 (2002) 244.


\bibitem{bib:sulvin}
J.D. Sullivian et al., Nucl. Instr. Meth. 95 (1971) 5.

\bibitem{bib:geant4}
S. Agostinelli et al., Nucl. Instr. Meth. A 506 (2003) 250.

\bibitem{bib:trop}
J. Bazo et al., ICRC (2013) 849.

\bibitem{bib:unfold}
A. Kondor, Nucl. Instr. Meth. 216 (1983) 177;
G. Agostini, Nucl. Instr. Meth. A 362 (1995) 487.

\bibitem{bib:bincen}
G.D. Lafferty, T.R. Wyatt, Nucl. Instr. Meth. A 355 (1995) 541.

\bibitem{bib:pamela}
O. Adriani, et al., Science, 332 (2011) 69;
O. Adriani, et al., Astrophys. J. 765 (2013) 91.
\bibitem{bib:ams-01}
J. Alcaraz et al., Phys. Lett. B 490 (2000) 27.
\bibitem{bib:atic}
A.D. Panov et al., Bull. Russian Acad. Sci. 73 (2009) 564.
\bibitem{bib:balloon}
M. Ichimura et al., Phys. Rev. D 48 (1993) 1949.
\bibitem{bib:bess-p}
Y. Shikaze et al., Astropart. Phys. 28 (2007) 154.
\bibitem{bib:bess-98}
T. Sanuki et al., Astrophys. J. 545 (2000) 1135.
\bibitem{bib:bess-tev}
S. Haino et al., Phys. Lett. B 594 (2004) 35.
\bibitem{bib:bess-pol}
K. Sakai et al., ICRC (2013) 974.
\bibitem{bib:capr-94}
M. Boezio et al., Astrophys. J. 518 (1999) 457.
\bibitem{bib:capr-98}
M. Boezio et al., Astropart. Phys. 19 (2003) 583.
\bibitem{bib:cream}
Y.S. Yoon et al., Astrophys. J. 728 (2011) 122.
\bibitem{bib:imax}
W. Menn et al., Astrophys. J. Lett. 533 (2000) 281.
\bibitem{bib:jacee}
K. Asakimori et al., Astrophys. J. 502 (1998) 278.
\bibitem{bib:mass-91}
R. Bellotti et al., Phys. Rev. D 60 (1999) 052002.
\bibitem{bib:rich-97}
E. Diehl et al., APh 18, 487 (2003)
\bibitem{bib:runjob}
M. Hareyama et al., J. Phys. Conf. 31 (2006) 159.
\bibitem{bib:sokol}
I.P. Ivanenko et al., Proc. ICRC 2 (1993) 17
\bibitem{bib:crdb}
D. Maurin et al., arxiv:1302.5525 (2013).


\end{thebibliography}
\end{document}